\documentclass[amsmath,amssymb]{revtex4}

\usepackage{graphicx}
\usepackage{dcolumn}
\usepackage{bm}
\usepackage{amsmath}
\usepackage{amssymb}

\newcommand{\be}{\begin{equation}}
\newcommand{\ee}{\end{equation}}
\newcommand{\sh}{2\pi \delta}
\newcommand{\sw}{2\pi\omega}
\newcommand{\fa}{\Psi_A}
\newcommand{\fb}{\Psi_B}
\newcommand{\fad}{\Psi_A^\dagger}
\newcommand{\fbd}{\Psi_B^\dagger}

\newcommand{\la}{\lambda}

\begin{document}

\title{Correlation Functions of One-Dimensional Lieb-Liniger Anyons}

\author{Ovidiu I. P\^{a}\c{t}u }
   \affiliation{C.N. Yang Institute for Theoretical Physics, State University
   of New York at Stony Brook, Stony Brook, NY 11794-3840, USA  }
   \altaffiliation{Permanent address: Institute of Space Sciences MG 23, 077125
   Bucharest-Magurele, Romania}
       \email{ipatu@grad.physics.sunysb.edu}

\author{Vladimir E. Korepin}
    \affiliation {C.N. Yang Institute  for Theoretical Physics, State University of
    New York at Stony Brook, Stony Brook, NY 11794-3840, USA}
       \email{korepin@max2.physics.sunysb.edu}

\author{Dmitri V. Averin}
    \affiliation{Department of Physics and Astronomy, State University of New York
    at Stony Brook, Stony Brook, NY 11794-3800, USA}
       \email{dmitri.averin@stonybrook.edu}

\begin{abstract}

We have investigated the properties of a model of 1D anyons
interacting through a $\delta$-function repulsive potential. The
structure of the quasi-periodic boundary conditions for the anyonic
field operators and the many-anyon wavefunctions is clarified. The
spectrum of the low-lying excitations including the particle-hole
excitations is calculated for periodic and twisted boundary
conditions. Using the ideas of the conformal field theory we obtain
the large-distance asymptotics of the density and field correlation
function at the critical temperature $T=0$ and at small finite
temperatures. Our expression for the field correlation function
extends the results in the literature obtained for harmonic quantum
anyonic fluids.

\end{abstract}

\maketitle

\section{Introduction}

For hard-core particles moving in two spatial dimensions, one can
unambiguously define the notion of braiding of the particle
trajectories by introducing the winding number $n$ that gives the
number of times the trajectory of one particle encircles another
particle. This fact makes it possible to consider ``anyonic''
particles with fractional exchange statistics \cite{LM,GMS}, for
which the wavefunction acquires the non-trivial phase factor $e^{\pm
i 2 \pi \kappa}$, where $\kappa$ is the ``statistical parameter'',
whenever $n$ changes by $\pm 1$. This situation can be contrasted
with the case of three spatial dimensions where one can define only
permutations (no braiding) of point-like particles leading to only
integer statistics, i.e. $\kappa=0,1$ for bosons and fermions,
respectively. In physical terms, the anyons in two dimensions can be
viewed as the charge-flux composites for which the statistical phase
arises as the result of the Aharonov-Bohm interaction between the
charge of one particle and the flux of the other \cite{FW}.
Experimentally, anyons can be realized as quasiparticles of the
two-dimensional (2D) electron liquids in the Fractional Quantum Hall
Effect (FQHE) \cite{ASW}. Individual quasiparticles are localized
and controlled by quantum antidots in the FQHE regime \cite{GS}, and
the transport properties of multi-antidot systems should provide
direct manifestations of their fractional exchange statistics
\cite{AN}. Dynamics of individual FQHE quasiparticles attracted
considerable attention (see, e.g., \cite{AG,SFN}) as a possible
basis for realization of the topological quantum computation
\cite{AK}.

Both conceptually and in practice (e.g., in FQHE systems), the 2D
anyons can be confined to move in one dimension. There are, however,
the aspects of fractional statistics in one dimension that make its
introduction more complicated than in two dimensions. One is that
for strictly 1D particles, a trajectory of one particle can not wind
around another, making the sign of the exchange phase $e^{\pm i \pi
\kappa sgn (x_i-x_j)/2}$ that the wavefunction should acquire when
the particle with coordinate $x_i$ moves past the one with $x_j$,
undetermined.  The sign of this phase depends on whether $x_i$
rotates clockwise or counter-clockwise around $x_j$ in the
underlying 2D geometry, which also explains why the signs of the
phase change at $x_i=x_j$ are opposite for the two particles in the
pair: rotation of one sense for increasing coordinate $x_i$ implies
the opposite rotation for increasing $x_j$. This fact hindered the
early attempts at direct introduction of the 1D anyons as
charge-flux composites \cite{SJBR,AGRPS}. It implies that any
description of the 1D anyons requires an additional convention on
the choice of the sign of the statistical phase for each pair of
particles. As discussed in more details below, this choice can be
arbitrary and affects the appropriate boundary conditions of the
quantum-mechanical wavefunctions of the system of anyons.

Another complicating aspect of the fractional statistics in 1D is
the interplay between the two types of statistics, exchange
statistics discussed in the preceding paragraph and the exclusion
statistics defined through the volume of the phase-space occupied by
one particle \cite{FDMH}. The exclusion statistics provides
effective description of the dynamic interaction of particles, while
the exchange statistics is associated with the ``real''
non-thermodynamic statistical effects that continue to exist in the
limit of hard-core particles with infinite repulsion. The model of
1D anyons with $\delta$-function interaction considered in this work
contains both types of effects, and the interplay between them can
be seen in Eq.~(\ref{c}) for the renormalization of the dynamic
particle-particle interaction by the exchange statistics. The
renormalized interaction determines the thermodynamics of the model
and can be expressed in terms of the exclusion statistics. However,
in the hard-core limit $c\rightarrow \infty$, effective interaction
constant is essentially independent of the exchange statistics, and
the main features of the thermodynamics of the model coincide with
that of free fermions. There are still anyonic effects (e.g., the
shift of the quasiparticle momenta by the parameter of the exchange
statistics $\kappa$) in this limit.

The purpose of this work is to provide a systematic description of
the ground state, low-lying excitations, and the asymptotics of the
correlation function of the gas of 1D anyons with the
$\delta$-function repulsion. In the form used below, the model was
introduced by Kundu \cite{Kundu}, who also provided the Bethe Ansatz
solution. It was further analyzed recently  by Batchelor {\em et
al.} \cite{BGO,BG,BGH}. The model is an anyonic extension of the
Bose gas with $\delta$-function interaction solved by Lieb and
Liniger \cite{LL}. The anyon gas reduces to the Bose gas in the
limit when the statistical parameter $\kappa$ vanishes. In general,
the Bethe equations for anyons are equivalent to the Bethe equations
for bosons with two effects of the statistics $\kappa$: renormalized
coupling constant, and a twist in the boundary conditions. This
makes it possible for us to use some of the known results for the
Bose gas in the discussion of anyons. (Detailed description of the
Bose gas with $\delta$-function interaction including the
correlation functions can be found in \cite{KBI}). The main results
of our work are the formulae (\ref{denscorgen}), (\ref{denscorgent})
for the density correlation function and (\ref{fieldcorgen}),
(\ref{fieldcorgent}) for the field correlation function. Expressions
for the field correlators extend the results of Calabrese and
Mintchev \cite{CM} obtained in the harmonic fluid approach by
including the higher-order terms that correspond to particle-hole
excitations. Also, the conformal field theory approach we use
provides immediate generalization of the zero-temperature
correlators to finite temperatures.

There are other 1D models of anyons in the literature. Liguori,
Mintchev and Pilo \cite{LMP} investigated the momentum distribution
of a more general gas of free anyons and predicted anyon
condensation in a certain range of the statistical parameter. Ilieva
and Thirring \cite{IT} studied the Hilbert space structure of the
anyonic field, and showed that for a fixed statistical parameter it
can be represented as an orthogonal sum of sectors with different
numbers of particles. The Hilbert space of our model has the same
structure.

The paper is organized as follows. Sections \ref{ftm} introduces the
field theoretical model of the 1D gas of anyons with periodic and
twisted boundary conditions. Equivalent quantum-mechanical problem
is formulated in Section \ref{qmep}. In Section \ref{pgs}, we
discuss the properties of the ground-state, and in Section \ref{fsc}
calculate the finite size corrections for the ground-state and
properties of the low-lying excitations. In Section \ref{cft}, using
the conformal field theory approach we find the large-distance
asymptotics for the zero-temperature density and field correlation
functions and correlators at small finite temperatures. Appendix
\ref{bcaa} presents the discussion of the boundary conditions for
many-anyonic wavefunctions used in Section \ref{qmep}. Appendix
\ref{pheappendix} describes the calculation of the energy and
momentum of particle-hole excitations for periodic and twisted
boundary conditions.

\section{The Lieb-Liniger Gas of Anyons}
\label{ftm}

We consider a gas of anyons with $\delta$-function interaction in
one dimension characterized by the Hamiltonian \be\label{hama}
H=\int_{0}^L dx \ \{ [\partial_x\fad(x)][\partial_x\fa(x)]
+c\fad(x)\fad(x)\fa(x)\fa(x) \} \, , \ee where $c>0$ is the coupling
constant and $L$ the length of the system. The anyonic fields obey
the equal-time commutation relations \be\label{com1}
\fa(x_1)\fad(x_2)=e^{-i\pi\kappa\epsilon(x_1-x_2)}\fad(x_2)\fa(x_1)+
\delta(x_1-x_2)\, , \ee \be\label{com2}
\fad(x_1)\fad(x_2)=e^{i\pi\kappa\epsilon(x_1-x_2)}\fad(x_2)\fad(x_1)\,
, \ee \be\label{com3}
\fa(x_1)\fa(x_2)=e^{i\pi\kappa\epsilon(x_1-x_2)}\fa(x_2)\fa(x_1)\, ,
\ee where \be \epsilon(x_1-x_2)=\left\{\begin{array}{rcr}
                          1 &\mbox{ when }& x_1>x_2\, , \\
                          -1&\mbox{ when }& x_1<x_2\, , \\
                          0& \mbox{ when }& x_1=x_2\, .
                         \end{array}\right.
\ee In the original work \cite{Kundu} introducing this model, the
anyonic fields were realized in terms of the bosonic fields
\be\label{def} \fad(x)=\fbd(x)e^{i\pi\kappa\int_{0}^x dx'\rho(x')},
\ \ \ \ \ \fa(x)=e^{-i\pi\kappa\int_{0}^x dx'\rho(x')}\fb(x)\, , \ee
where \be \rho(x)\equiv\fad(x)\fa(x)=\fbd(x)\fb(x) \, . \ee Due to
the fact that at coinciding points $\epsilon(0)=0$, the commutation
relations (\ref{com1},\ref{com2},\ref{com3}) are indeed bosonic. An
alternative realization in terms of the fermionic fields was
proposed in \cite{Gir}. However, in this case, the interaction term
in the Hamiltonian (\ref{hama}) vanishes, since
$\fa^2(x)=[\fad(x)]^2=0$ in coinciding points (see also the
discussion in \cite{BGH}). One implication of this difference is
that in comparison to the bosonic representation (\ref{def}),
similar fermionic representation with appropriate modification of
the statistical parameter, effectively makes it possible to describe
only the infinite repulsion limit $c \rightarrow \infty$.

Characteristics of the anyonic gas (\ref{hama}) depend on the
boundary conditions imposed on the system at $x=0=L$. In this work,
we use two different quasiperiodic boundary conditions which impose
periodicity either directly on the anyonic or on the bosonic fields.
Equations (\ref{def}) imply that the periodic boundary condition for
anyons correspond to twisted boundary conditions for bosons and
viceversa. In terms of the anyonic fields, the boundary condition we
use are, \be \mbox{periodic BC:} \ \ \ \ \fad(0)=\fad(L)\, , \ee and
\be \label{tw} \mbox{twisted BC:} \ \ \ \ \fad(0)=\fad(L )e^{-i\pi
\kappa (N-1)} \, , \ee where $N$ is the number of particle in the
system. One can see directly from Eq.~(\ref{def}) that the external
phase shift $\pi \kappa (N-1)$ introduced into the conditions
(\ref{tw}), ensures the periodicity of the bosonic fields. As will
be shown in more details below, this means that this phase removes
the anyonic shift of the quasiparticle momenta. Below, we use the
common notation for the two types of boundary conditions: \be
\label{cbc} \fad(0)=\fad(L)e^{-i\pi \beta \kappa (N-1)}\, , \ \ \
\beta=0,1\, . \ee An important difference of the anyons with
fractional exchange statistics from the integer-statistics particles
is that the boundary conditions (\ref{cbc}) for the fields do not
translate directly into the same boundary conditions for the
quantum-mechanical wavefunctions of the $N$-anyon system \cite{AN},
which have more complicated structure (\ref{boundarycond}) derived
in the Appendix \ref{bcaa}.

The corresponding equation of motion
$-i\partial_t\fa(x,t)=[H,\fa(x,t)]$ for the boundary conditions
(\ref{cbc}) is the nonlinear Schr\"odinger equation \be
i\partial_t\fa(x,t)=\partial_x\fa(x,t)+2c\fad(x,t)\fa^2(x,t) \, .\ee
The number of particle operator $Q$ and the momentum operator $P$
are defined as \be Q=\int_0^Ldx\ \fad(x)\fa(x)\, , \ee \be
P=-\frac{i}{2}\int_0^Ldx\
\left(\fad(x)\partial_x\fa(x)-[\partial_x\fad(x)]\fa(x)\right)\, .
\ee Both of them are hermitian operators which commute with the
Hamiltonian \be [H,P]=[H,Q]=0 \, . \ee If we define the Fock vacuum
as \be \fa(x)|0\rangle\ \, , \;\; x\in[0,L] \, , \ee the
$N$-particle eigenstate of the Hamiltonian (and also of $P$ and $Q$)
can be then written as \be |\psi\rangle_N=\int d^Nx \
e^{-\frac{i\pi\kappa
N}{2}}\chi_N(x_1,\cdots,x_N)\fad(x_1)\cdots\fad(x_N)|0\rangle \, ,
\ee where the many-body wavefunction obeys
\be\label{anyonicproperty}
\chi_N(x_1,\cdots,x_i,x_{i+1},\cdots,x_N)=e^{-i\pi\kappa \epsilon
(x_i-x_{i+1})}\chi_N(x_1,\cdots,x_{i+1},x_{i},\cdots,x_N) \, . \ee
This can be seen directly by using the exchange relation of the
field operators $\fad(x_i)\fad(x_{i+1})=e^{i\pi \kappa \epsilon
(x_i-x_{i+1})} \fad(x_{i+1})\fad(x_{i})$ and interchanging the name
of the integration variables $x_i$, $x_{i+1}$. Iterating the
exchanges several times we obtain \be
\chi_N(x_1,\cdots,x_i,\cdots,x_{j},\cdots,x_N)=e^{-i\pi \kappa
[\sum_{k=i+1}^j \epsilon(x_i-x_{k}) -\sum_{k=i+1}^{j-1}\epsilon
(x_j-x_k)]} \chi_N(x_1,\cdots,x_{j},\cdots,x_{i},\cdots,x_N)\, . \ee

\section{ The Equivalent Quantum Mechanical Problem}
\label{qmep}

In \cite{Kundu,BGH}, it was shown that the eigenvalue problem (for
periodic boundary conditions) \be H|\psi\rangle_N = E_N |\psi
\rangle_N  ,\;\;\;\;\; P|\psi\rangle_N=p_N|\psi\rangle_N  ,
\ \ 
\ee can be reduced to the quantum-mechanical problem \be \mathcal{H}
|\psi\rangle_N= E_N|\psi\rangle_N  , \;\;\;\;\; \mathcal{P} |\psi
\rangle_N=p_N|\psi\rangle_N  ,
\ee where \be \label{hamq} \mathcal{H}_N =\sum_{j=1}^N \left(-\frac{
\partial^2}{\partial x_j^2}\right)+2c\ \sum_{1\leq j\leq k \leq N}
\delta(x_j-x_k)\, , \ee \be
\mathcal{P}=\sum_{j=1}^N\left(-\frac{\partial}{\partial_{x_j}}\right)
\, . \ee These considerations also hold for twisted and all cyclic
boundary conditions for field operators. The boundary conditions for
the quantum-mechanical wavefunctions of $N$ anyons are (see
\cite{AN} and Appendix \ref{bcaa})
\begin{eqnarray} \chi_N(0,x_2,\cdots,x_N)& =&
e^{i\pi\beta\kappa(N-1)}\ \ \ \ \ \ \ \ \ \chi_N(L,x_2,\cdots,x_N)
\, , \nonumber
\\ \chi_N(x_1,0,\cdots,x_N) &=&e^{-i2\pi\kappa}e^{i\pi\beta\kappa(N-1)}
\chi_N(x_1,L,\cdots,x_N) \, , \nonumber \\
                 & \vdots & \label{boundarycond}  \\
\chi_N(x_1,x_2,\cdots,0)&=&e^{-i2(N-1) \pi\kappa} e^{i\pi\beta\kappa
(N-1)} \chi_N(x_1,x_2\cdots,L) \, , \nonumber \end{eqnarray} where,
as defined above, $\beta=0,1$ for periodic and twisted boundary
conditions (\ref{cbc}).

Using the Coordinate Bethe Ansatz \cite{Kundu,BGO,BGH} we can obtain
the eigenfunctions of the Hamiltonian (\ref{hamq}) as \be \label{wf}
\chi_N=\frac{e^{-i\frac{\pi\kappa}{2}\sum_{j<k}\epsilon(x_j-x_k)}}{
\sqrt{N!\prod_{j>k}[(\lambda_j-\lambda_k)^2+c'^2]}}
\sum_{\mathcal{P}}(-1)^{[\mathcal{P}]}e^{i\sum_{n=1}^Nx_n
\lambda_{\mathcal{P}_n}} \prod_{j>k}[\lambda_{\mathcal{P}_j}
-\lambda_{\mathcal{P}_k}-ic'\epsilon(x_j-x_k)] \, , \ee where
$[\mathcal{P}]$ is the signature of the permutation and \be
c'=\frac{c}{\cos(\pi\kappa/2)} \label{c} \ee is the coupling
constant renormalized by the exchange statistics. The eigenvalues of
the Hamiltonian and momentum operators are $E_N=\sum_{j=1}^N\la_j^2$
and $p_N=\sum_{j=1}^N\la_j$, respectively. For the boundary
conditions (\ref{boundarycond}) we obtain the Bethe equations \be
\label{BE} e^{i\lambda_jL}=e^{i\pi(1-\beta)\kappa(N-1)}\prod_{k=1,k
\ne j}^N\left(\frac{\lambda_j-\lambda_k+ic'}{\lambda_j-\lambda_k-ic'
} \right) . \ee The Bethe equations (\ref{BE}) are similar to those
obtained by Lieb and Liniger for the Bose gas with repulsive
$\delta$-function interaction. In our case, however, the effective
coupling constant $c'$ (\ref{c}) can take negative values. While it
can be shown (see, e.g., \cite{KBI}) that the Bethe roots $\la_j$
are real for $c'>0$ , the roots can become complex for $c'<0$, and
one gets bound states \cite{Mcguire}. In this work, we will consider
only the case $c'>0$.

\section{Properties of the Ground State}
\label{pgs}

Bethe equations (\ref{BE}) can also be written as \be \label{BEM}
\lambda_jL +\sum_{k=1}^N\theta(\lambda_j-\lambda_k)=2\pi n_j+ \pi
\kappa (1-\beta)(N-1)\, , \ \ \ \ \ j=1,\cdots,N \, , \ee where \be
\theta(\lambda)=i\ln\left(\frac{ic'+\lambda}{ic'-\lambda}\right)\, ,
\ee and $n_j$ are integers when $N$ is odd and half-integers when
$N$ is even.

\subsection{Twisted Boundary Conditions }

In this case ($\beta=1$), the Bethe equations are similar to those
for the Bose gas with periodic boundary conditions \cite{LL,KBI}
with $c'$ as a coupling constant. The ground state is characterized
by the set of integers (half-integers) $n_j=j-(N+1)/2$, so the Bethe
equations take the form \be\label{gstbc} \lambda_j^BL+\sum_{k=1}^N
\theta(\lambda_j^B-\lambda_k^B)=2\pi \left(j-\frac{N+1}{2}\right)\,
, \ \ \ \ j=1,\cdots,N \, . \ee From now on the superscript $B$ will
mean that the variables and physical quantities are the same as the
ones for the Bose gas with periodic boundary conditions and coupling
constant $c'$. In the thermodynamic limit $N,L\rightarrow\infty, \,
D=N/L =$const, the Bethe roots become dense and fill the symmetric
interval $[-q,q]$.  The density of roots in this interval obeys the
Lieb-Liniger integral equation \be
\rho(\la)-\frac{1}{2\pi}\int_{-q}^qK(\lambda,\mu)\rho(\mu)\
d\mu=\frac{1}{2\pi} \, , \ee where $
K(\la,\mu)=\theta'(\la-\mu)=2c'/(c'^2+(\la-\mu)^2)$. The Fermi
momentum $q$ can be obtained from the Lieb-Liniger integral equation
and the particle density is \be D=\frac{N}{L}=\int_{-q}^q\rho(\la)\
d\la \, . \ee Finally, the energy and the momentum of the ground
state are \be\label{energytbc} E^B_0=L\int_{-q}^q\la^2\rho(\la)\
d\la\, , \ \ \ \ \ \ P^B_0=0\, . \ee

\subsection{Periodic Boundary Conditions }

This is the case treated in \cite{BGO,BG,BGH}. The Bethe equations
(\ref{BEM}) in this case ($\beta=0$) are similar to those for the
Bose gas with twisted boundary conditions: \be
\lambda_jL+\sum_{k=1}^N \theta(\lambda_j -\lambda_k)=2\pi
n_j+\pi\kappa(N-1)\, , \ \ \ \ \ j=1,\cdots,N \, . \ee Introducing
the notation $\{[...]\}$ such that \be\label{fractionalpart}
\{[x]\}=\gamma\, , \ \ \ \ \mbox{ if } x=2\pi \times \mbox{ integer
}+2\pi\gamma \, , \ \ \ \gamma\in[0,1) \, , \ee we can describe the
ground state by the following set of the Bethe equations:
\be\label{gspbc} \lambda_jL+\sum_{k=1}^N \theta(\lambda_j
-\lambda_k)=2\pi\left(j-\frac{N+1}{2}\right)+\sh \, , \ \ \ \
j=1,\cdots,N \, , \ee where $\delta=\{[\pi\kappa(N-1)]\}$.
Comparison of Eqs.~(\ref{gspbc}) and (\ref{gstbc}) shows that we
have the following connection between the Bethe roots for periodic
and twisted boundary conditions: \be \label{del} \lambda_j=
\lambda_j^B +\sh/L \, . \ee This relation is exact and holds also
for the excited states if the (half)integers in the Bethe equations
are the same. In the periodic case, the ground state is shifted by
$\sh/L$, so that the Bethe roots are now distributed in the interval
$[-q+\sh/L,q+\sh/L]$, and momentum of the ground state $P_0$ in
general does not vanish: \be \label{P0}
P_0=\sum_{i=1}^N\lambda_i=\sum_{i=1}^N(\lambda_i^B+2\pi\delta/L)=2\pi
D\delta \, . \ee The ground-state energy is: \be \label{energypbc}
E_0 = \sum_{i=1}^N \lambda_i^2=\sum_{i=1}^N\left((\lambda_i^B)^2+
\frac{4\pi\delta\lambda_i^B}{L}+\frac{(2\pi\delta)^2}{L^2}\right) =
E^B_0+\frac{D(2\pi\delta)^2}{L} \, , \ee where we have used that the
total momentum in the case of twisted boundary conditions is zero
and $E^B_0$ in the thermodynamic limit is given by
Eq.~(\ref{energytbc}).

\section{Finite Size Corrections}
\label{fsc}

In this section, we are going to calculate the finite size
corrections for the energy of the ground state and characteristics
of the low-lying excitations. Based on the results of this section,
we will be able to find  the large-distance asymptotics of
the correlations functions using conformal field theory. A chemical
potential $h$ is added to the Hamiltonian (\ref{hama}) throughout
this section, so that the total Hamiltonian is \be H_h=\int_{0}^L dx
\ \{ [\partial_x\fad(x)] [\partial_x\fa(x)] + c\fad(x)
\fad(x)\fa(x)\fa(x) -h\fad(x)\fa(x) \} \, . \ee

\subsection{Finite Size Corrections for The Ground State Energy}

As we have seen in the previous section, the ground state of the gas
of anyons with twisted boundary conditions ($\beta=1$) is
characterized by the same set of Bethe equations as the Bose gas
with coupling constant $c'$ and periodic boundary conditions. So in
this case we can use the results for the Bose gas
\cite{KBI,BIR,BIR1,BM1,BM2,WET}: \be\label{gsenergy}
E^B_0=L\int_{-q}^q\varepsilon_0(\la)\rho(\la)\ d\la-\frac{\pi
v_F}{6L}+\mathcal{O}\left(\frac{1}{L^2}\right) , \ee where
$\varepsilon_0(\la)=\la^2-h$ and $v_F$ is the Fermi velocity for the
Bose gas with coupling constant $c'$. In the case of periodic
boundary conditions ($\beta=0$), Eq.~(\ref{energypbc}) then gives:
\be E_0 = L\int_{-q}^q\varepsilon_0(\la)\rho(\la)\ d\la-\frac{\pi
v_F}{6L}+\frac{D(2\pi\delta)^2}{L}+\mathcal{O}\left(\frac{1}{L^2}
\right) .\\ \ee

\subsection{Low-Lying Excitations}

In our discussion of the low-lying excitations, we consider several
different types of excitation processes:
\begin{itemize}
\item Addition of a finite number $\Delta N$ of particles into the
ground state of the system.
\item Backscattering: all integers $n_j$ in the set $\{n_j\}$
characterizing the ground-state distribution are shifted by an
integer $d$.
\item Particle-hole excitations: the integer $n_j$ that
characterizes the particle at the Fermi surface is  modified from
its value in the ground state distribution by $N^+$ for the particle
with momentum $q$, (or $q+\sh/L$, depending on the boundary
conditions) or by $N^-$ at the opposite point of the Fermi surface
with momentum $-q,(-q+\sh/L)$.
\end{itemize}
The central feature of the gas of anyons is that the boundary
conditions for the field operators and the wavefunctions depend on
the number of particles in the system. This means that any
modification of the number of particles in the system changes the
Bethe equations and, as a result, the quasiparticle momenta given by
the Bethe roots. If we add one particle to the system of $N$
particles, the boundary conditions are:
\begin{eqnarray}
\chi_{N+1}(0,x_2,\cdots,x_N,x_{N+1})&=&e^{i\pi\beta\kappa(N-1)}
\chi_{N+1}(L,x_2,\cdots,x_N,x_{N+1})\, , \nonumber\\
\chi_{N+1}(x_1,0,\cdots,x_N,x_{N+1})&=&e^{-i2\pi\kappa}
e^{i\pi\beta\kappa(N-1)}\chi_{N+1}(x_1,L,\cdots,x_N,x_{N+1})\, , \nonumber\\
                 &\vdots& \\
\chi_{N+1}(x_1,x_2,\cdots,0,x_{N+1})&=&e^{-i2(N-1)\pi\kappa}
e^{i\pi\beta\kappa(N-1)}\chi_{N+1}(x_1,x_2\cdots,L,x_{N+1})\, , \nonumber\\
\chi_{N+1}(x_1,x_2,\cdots,x_N,0)\ \ \
&=&e^{-i2N\pi\kappa}e^{i\pi\beta\kappa(N-1)}\chi_{N+1}(x_1,x_2\cdots,x_{N},L)\,
, \nonumber \end{eqnarray} and the Bethe equations become \be
e^{i\lambda_jL}=e^{i\pi\kappa N}e^{-i\pi\beta\kappa(N-1)} \prod_{
k=1,k \ne j}^{N+1}\left(\frac{\lambda_j- \lambda_k+ic'}{ \lambda_j
-\lambda_k-ic'}\right). \ee The ground states for $N$ and $N+1$
particles are characterized by the Bethe roots satisfying different
equations:
\begin{eqnarray}\label{eqn1}
\lambda_jL+\sum_{k=1}^N\theta(\lambda_j-\lambda_k)&=&2\pi\left(j-
\frac{N+1}{2}\right)+2\pi\omega\, ,  \ \ \ \ \ j=1,\cdots,N \, , \nonumber\\
\tilde\lambda_jL+\sum_{k=1}^{N+1}\theta(\tilde\lambda_j-\tilde\lambda_k)
&=&2\pi\left(j-\frac{N+2}{2}\right)+2\pi\omega', \ \ \ \ \
j=1,\cdots,N+1 \, ,
\end{eqnarray}
where
\[ \omega=0, \;\; \omega'=\kappa/2\, , \;\;\; \mbox{and} \;\;\;
\omega=\{[\pi\kappa(N-1)]\}, \;\;  \omega'=\{[\pi\kappa N]\}, \]
for the twisted ($\beta=1$) and periodic ($\beta=0$) boundary
conditions, respectively, and $\{[...]\}$ is defined by
Eq.~(\ref{fractionalpart}).
Comparing Eq.~(\ref{eqn1}) with Eq.~(\ref{gstbc}) we see that
\be\label{rel} \la_j=\la_{jN}^B+\sw/L\, , \ \ \ \ \ \
\tilde\lambda_j=\la_{j,N+1}^B +\sw'/L \, , \ee where $\la_{jN}^B$
are the Bethe roots characterizing the ground state of a gas of $N$
bosons with periodic boundary conditions and coupling constant $c'$.

\subsubsection{Addition of One Particle to the System}

For excitations of this type we assume that both before and after
the addition of a particle, the system is in the ground state. In
order to calculate the energy and momentum of this excitation, we
use Eq.~(\ref{rel}) which enables one to express energy and momentum
through corrections to the same characteristics of excitations of
the Bose gas.

For the energy we get from Eq.~(\ref{rel}):
\begin{eqnarray}\label{var1}
\Delta E(\Delta N=1)&=&\sum_{j=1}^{N+1}\varepsilon_0
(\tilde\lambda_j)-\sum_{j=1}^{N}\varepsilon_0(\lambda_j)
\nonumber\\
&=&\Delta E^B(\Delta N=1)+(N+1)\left(\frac{\sw'}{L} \right)^2
-N\left(\frac{\sw}{L}\right)^2\, ,
\end{eqnarray}
where $\Delta E^B(\Delta N=1)$ is the energy of the corresponding
bosonic excitation. As known in the literature (see, e.g.,
\cite{KBI,BIR,BIR1,BM2,WET}) it is convenient to express this energy
in terms of the "dressed charge" $Z(\la)$: \be\label{var2} \Delta
E^B(\Delta N=1)=\frac{2\pi v_F}{L} \left(
\frac{1}{2\mathcal{Z}}\right)^2, \ee where $\mathcal{Z}=Z(q)=Z(-q)$,
and $Z(\la)$ is defined as solution of the equation \be\label{zie}
Z(\la)-\frac{1}{2\pi}\int_{-q}^q K(\la,\mu)Z(\mu)\ d\mu=1 \, . \ee
From (\ref{var1}) and (\ref{var2}) we obtain \be\label{1penergy}
\Delta E(\Delta N=1)=\frac{2\pi v_F}{L}\left(\frac{1}{2\mathcal{Z}}
\right)^2 +(N+1) \left(\frac{\sw'}{L}\right)^2-N\left( \frac{
\sw}{L}\right)^2 . \ee

The momentum of the excitation is: \be \label{1pmomenta} \Delta
P(\Delta N=1)=\sum_{j=1}^{N+1}\tilde\lambda_j-\sum_{j=1}^{N}
\lambda_j = (N+1)\frac{\sw'}{L} -N \frac{\sw}{L} \, , \ee where we
again used the fact that for the ground state of bosons with
periodic boundary conditions and any number of particles the total
momentum is vanishing.

\subsubsection{Backscattering}

The uniform shift of the ground-state distribution in a
backscattering process can be understood as a jump of some number
$d$ of particles between the opposite boundaries of the Fermi
surface. The Bethe equations relevant for this process (in the case
of $N$ and $N+1$ particles in the ground state) take the form:
\begin{eqnarray}
\lambda_j^dL+\sum_{k=1}^N\theta(\lambda_j^d-\lambda_k^d)&=&2\pi
\left(j-\frac{N+1}{2}\right)+2\pi d+2\pi\omega\, , \ \ \ \ \
j=1,\cdots,N \, , \nonumber\\
\tilde\lambda_j^dL+\sum_{k=1}^{N+1}\theta(\tilde\lambda_j^d-
\tilde\lambda_k^d)&=&2\pi\left(j-\frac{N+2}{2}\right)+2\pi
d+2\pi\omega',\ \ \ \ \ j=1,\cdots,N+1\, .
\end{eqnarray}
Again, comparison with Eq.~(\ref{gstbc}) shows that \be \label{rel1}
\la_j^d=\la_{jN}^B+2\pi(\omega+d)/L\, , \ \ \ \ \
\tilde\lambda_j^d=\la_{j,N+1}^B+2\pi(\omega'+d)/L \, ,\ee and the
ground states are characterized by Eq.~(\ref{rel}). Using
Eqs.~(\ref{rel}) and (\ref{rel1}) we get the excitation energy:
\begin{eqnarray}
N \mbox{ particles: }\ \ \ \ \Delta E(d)&=&\sum_{j=1}^N(
\varepsilon_0 (\la_j^d) -\varepsilon_0(\la_j))
=N\frac{(2\pi\omega+2\pi d)^2}{L^2}-N\frac{(2\pi\omega)^2}{L^2}\, , \\
N+1 \mbox{ particles: }\ \ \ \ \Delta
E(d)&=&\sum_{j=1}^{N+1}(\varepsilon_0(\tilde\la_j^d)-\varepsilon_0(\tilde\la_j))
=(N+1)\frac{(2\pi\omega'+2\pi
d)^2}{L^2}-(N+1)\frac{(2\pi\omega')^2}{L^2}\, .
\end{eqnarray}
This result can be rewritten using the relation $\mathcal{Z}^2=2\pi
D/v_F$ (see \cite{KBI}, Chap.~I.9) obtaining
\begin{eqnarray}\label{backnenergy}
N \mbox{ particles: }\ \ \ \ \Delta E(d)&=&\frac{2\pi v_F}{L}\mathcal{Z}^2
(d+\omega)^2-\frac{2\pi v_F}{L}\mathcal{Z}^2\omega^2, \nonumber\\
N+1 \mbox{ particles: }\ \ \ \ \Delta E(d)&=&\frac{2\pi
v_F}{L}\mathcal{Z}^2(d+\omega')^2-\frac{2\pi
v_F}{L}\mathcal{Z}^2\omega'^2+\frac{(2\pi\omega'+2\pi
d)^2}{L^2}-\frac{(2\pi\omega')^2}{L^2} \, .
\end{eqnarray}
The momentum of the backscattering excitation is simply \be
\label{backnmomenta} \Delta P(d)=N(2\pi d/L)\, , \ee the expression
that is valid for any number of particles $N$.

\subsubsection{Particle-Hole Excitations at the Fermi Surface}

In this case, the excitations we consider consist in changing the
maximal (minimal) $n_j$ in the ground state by $N^\pm$. For $N$
particles and ``excitation magnitude'' $N^+$ the Bethe equations are

\begin{eqnarray} \label{eq59}
\lambda_jL+\sum_{k=1}^N\theta(\lambda_j-\lambda_k)&=&2\pi
\left(j-\frac{N+1}{2}\right)+2\pi\omega\, , \ \ \ \ \
 j=1,\cdots,N-1 \, ,\nonumber\\
\la_N L+\sum_{k=1}^N\theta(\lambda_N-\lambda_k)&=&2\pi
\left(N-\frac{N+1}{2}\right)+2\pi\omega+2\pi N^+ .
\end{eqnarray}

From (\ref{eq59}) we see that the momentum of the excitation $N^+$
is $\Delta P(N^+)=2\pi N^+/L$ and, similarly, for the excitation
$N^-$ the momentum is $\Delta P(N^-)=-2\pi N^- /L$. These excitation
can be considered as a special case of the general particle-hole
excitations, and we can use the results of Appendix
\ref{pheappendix} for them. Using (\ref{phem}) we see that the
excitation energy and momentum \be\label{jumpn} \Delta
E(N^\pm)=\frac{2\pi v_F}{L}N^\pm +\mathcal{O} \left( \frac{1}{
L^2}\right) \, , \ \ \ \ \ \Delta P(N^\pm)=\pm\frac{2\pi}{L}N^\pm ,
\ee coincide with those for the similar excitations of the Bose gas
(see Appendix I.4 of \cite{KBI}): \be \Delta E^B(N^\pm)=\frac{2\pi
v_F}{L}N^\pm, \ \ \ \ \ \ \Delta P^B(N^\pm)=\pm\frac{2\pi}{L}N^\pm .
\ee  For $N+1$ particles, the energy and momentum of the excitations
are given by the same expressions as in (\ref{jumpn}).

\section{Large-Distance Asymptotics of Correlations Functions}
\label{cft}

In this section we calculate the asymptotics of the correlation
functions. We will consider the case of twisted boundary conditions
($\beta=1$), or the periodic boundary conditions ($\beta=0$) when
$\kappa$ is a integer multiple of $2/(N-1)$, so that the shift in
(\ref{del}) vanishes, $\delta=0$, and the two boundary conditions
are equivalent -- see (\ref{cbc}). The main feature of this case
that is important for the direct applicability of the conformal
field theory approach is that the momentum of the ground state
(\ref{P0}) of the gas of anyons is zero for these boundary
conditions. For general gapless 1+1-dimensional systems, $T=0$ is a
critical point making the correlation functions decay as a power of
distance at $T=0$ but exponentially at $T>0$. As we have seen in the
previous section, the Lieb-Liniger anyonic gas is gapless and the
excitation spectrum has a linear dispersion law in the vicinity of
the Fermi level. These features support the expectation that the
critical behavior of the anyon system is described by conformal
field theory (CFT).

CFT is a vast subject and we refer the reader to
\cite{BPZ,ISZ,Cardy1,Ginsparg} and Chap. XVIII of \cite{KBI} for
more information. A conformal theory is characterized by the central
charge $c$ (not to be confused with the coupling constant in
(\ref{hama})) of the underlying Virasoro algebra, and conformal
invariance constrains the critical behavior of the systems under
consideration. The critical exponents (the powers that characterizes
the algebraic decay at $T=0$) are related to the conformal
dimensions of the operators within the CFT, so to obtain the
complete information about the critical behavior of the system we
need to calculate the central charge and the conformal dimensions of
the primary fields.

\subsection{Central Charge}

In order to find the central charge we use the fact that for unitary
conformal theories it can be found from the finite-size corrections,
specifically the coefficient of the $1/L$ term in the expansion of
the ground state energy for $L\rightarrow\infty$ \cite{BCN,Affleck}:
\be E=L\epsilon_\infty-\frac{\pi v_F}{6L}c +\mathcal{O}\left(
\frac{1}{L}\right) \, . \ee Comparing this relation to
Eq.~(\ref{gsenergy}) valid for the boundary conditions we are
assuming in this Section, we see that the central charge $c=1$. The
fact that the central charge $c=1$ means that the critical exponents
can depend continuously on the parameters of the model
\cite{BPZ,FQS,DVV}.

\subsection{Conformal Dimensions from Finite Size Effects}

Following the original idea of Cardy \cite{Cardy2} subsequently
developed in \cite{BIR,BIR1,BM2}, we obtain below the conformal
dimensions of the conformal fields in the theory from the spectrum
of the low-lying excitations described in the previous Section. The
local fields of the model can be represented as a combination of
conformal fields \be\label{expansion}
\phi(x,t)=\sum_Q\tilde{\tilde{A}}(Q)\phi_Q(z,\bar{z})\, , \ee where
$\tilde{\tilde{A}}(Q)$ are some coefficients and $z=ix+v_F\tau$,
with $v_F$ the Fermi velocity and $\tau$ the Euclidean time. The
conformal fields are related to excitations with quantum numbers
$Q=\{\Delta N,N^\pm,d\}$, where $\Delta N$ represents the number of
particles created by the field $\phi$, and all the fields in the
expansion (\ref{expansion}) should have the same $\Delta N$. The
quantum number $d$ gives the number of particles backscattered
across the Fermi ``sphere'', and $N^\pm$ characterizes the change of
the maximal or minimal $n_j$ in the Bethe equations from its values
in the ground state. While $\Delta N$ has to be the same for all the
terms in the expansion, $d$ and $N^\pm$ can be different.

For two conformal fields, $\phi_Q$ and $\phi_{Q'}$,  with the same
conformal dimensions denoted $\Delta^\pm$, their correlation
function is given by \be\label{correlation}
\langle\phi_Q(z_1,\bar{z}_1)\phi_{Q'}(z_2,\bar{z}_2)\rangle=
\frac{1}{(z_1-z_2)^{2\Delta^+}(\bar{z}_1-\bar{z}_2)^{2\Delta^-}}.
\ee Under a conformal transformation
$z=z(w),\bar{z}=\bar{z}(\bar{w})$, it transforms like
\be\label{transform}
\langle\phi_Q(w_1,\bar{w}_1)\phi_{Q'}(w_2,\bar{w}_2)\rangle=
\prod_{j=1}^2\left(\frac{\partial z_j}{\partial
w_j}\right)^{\Delta^+}\left(\frac{\partial \bar{z}_j}{\partial
\bar{w}_j}\right)^{\Delta^-}
\langle\phi_Q(z_1(w_1),\bar{z}_1(\bar{w}_1))\phi_{Q'}(z_2(w_2),
\bar{z}_2(\bar{w}_2))\rangle\, . \ee

Using the expansion (\ref{expansion}), the fact that the two
conformal fields with different conformal dimensions are orthogonal
(their correlation function is zero), and (\ref{correlation}) we
then have: \be\label{corrinter} \langle
\phi(z_1,\bar{z}_1)\phi(z_2,\bar{z}_2)\rangle=\sum_Q\frac{\tilde{A}(Q)
}{(z_1-z_2)^{2\Delta^+_Q}(\bar{z}_1-\bar{z}_2)^{2\Delta^-_Q}}\, ,
\ee which is valid in the whole complex plane without the origin
($z_1\ne z_2$). Conformal mapping of this plane to a cylinder
(periodic strip) with the help of transformation \be\label{confmap}
z=e^{2\pi w/L}\, ,\ \ w=ix+v_F\tau  \;\; \mbox{ with } \;\; 0<x\leq
L\, , \ee applied to (\ref{transform}) gives \be \langle
\phi(w_1,\bar{w}_1)\phi(w_2,\bar{w}_2)\rangle=\sum_Q\tilde{A}(Q)
\left(\frac{\pi/L}{\sinh[\pi(w_1-w_2)/L]}\right)^{2\Delta^+_Q}
\left(\frac{\pi/L}{\sinh[\pi(\bar{w}_1-\bar{w}_2)/L]}\right)^{2\Delta^-_Q}
, \ee with the asymptotics \be \langle
\phi(w_1,\bar{w}_1)\phi(w_2,\bar{w}_2)\rangle\sim\sum_Qe^{-\frac{2\pi
v_F}{L}(\Delta^+_Q+\Delta^-_Q)(\tau_1-\tau_2)
-i\frac{2\pi}{L}(\Delta^+_Q-\Delta^-_Q)(x_1-x_2)}\, . \ee Comparison
with the spectral decomposition of the correlation function in the
periodic strip ($\tau_1>\tau_2$) \be \langle
\phi(w_1,\bar{w}_1)\phi(w_2,\bar{w}_2)\rangle_L=\sum_Q|\langle
0|\phi(0,0)|Q\rangle|^2e^{-(E_Q-E_0)(\tau_1-\tau_2)-i(P_Q-P_0)(x_1-x_2)}\,
, \ee where $|0\rangle$ is the ground state and $E_0\, ,P_0$ are the
energy and momentum of the ground state, leads to
\be\label{gapcdconn} E_Q-E_0=\frac{2\pi
v_F}{L}(\Delta^+_Q+\Delta^-_Q)\, , \ \
P_Q-P_0=\frac{2\pi}{L}(\Delta^+_Q-\Delta^-_Q)\, , \ee assuming that
both the energy and momentum gaps are of order $\mathcal{O}(1/L).$
However, as we have seen in Sect.~\ref{fsc}, for some of the
excitations considered (addition of a particle in the system,
$\Delta N=1$, backscattering processes characterized by $d$, and
particle-hole excitations at the Fermi surface characterized by
$N^\pm$), the momentum gap is macroscopic. For example, if
$Q=\{\Delta N=0,d\ne 0,N^\pm=0\} $, the momentum gap is $2k_Fd$,
$k_F\equiv \pi D$, and for $Q=\{\Delta N=1,d=0,N^\pm=0\}$ the
momentum gap is $\pi k_F \kappa+\pi\kappa/L$. For these excitations,
following \cite{BIR,BIR1,BM2}, the coefficients $\tilde{A}(Q)$ will
depend on $x$ as \be\label{xcoefdepen} \tilde{A}(Q)=A(Q)e^{ip_Qx}\,
, \ee where $p_Q$ is the macroscopic part of the momentum gap
$P_Q-P_0$. From (\ref{corrinter}) and (\ref{xcoefdepen}) we obtain
the generic formula for the asymptotics of correlations functions at
$T=0$ \be\label{corrgeneral} \langle
\phi(x,t)\phi(0,0)\rangle=\sum_Q\frac{A(Q)e^{ip_Qx}}{(ix+v_F\tau)^{2
\Delta^+_Q}(-ix+v_F\tau)^{2\Delta^-_Q}}\, , \ee where $\Delta^\pm_Q$
can be found from (\ref{gapcdconn}) and the leading term corresponds
to the smallest $\Delta^\pm_Q$.

We also can find the low-temperature asymptotics of the correlation
functions if we use instead of the conformal mapping
(\ref{confmap}), the mapping \be \label{tmap} z=e^{2\pi T w/v_F}\,
,\ \ \ z=x-iv_F\tau\, , \ee which differ from (\ref{confmap}) by
interchanging the space and time variables. The computations are
similar those described above for the correlation functions in a
finite box, and the final result is \be\label{cortemp} \langle
\phi(x,t)\phi(0,0)\rangle_T=\sum_Q B(Q)e^{ip_Qx}\left(\frac{\pi
T/v_F}{\sinh[\pi T(x-iv_F\tau)/v_F]}\right)^{2\Delta^+_Q}
\left(\frac{\pi T/v_F}{\sinh[\pi T(x+iv_F\tau)/v_F]}
\right)^{2\Delta^-_Q}\, . \ee This result is valid only at
temperatures close to zero.

\subsection{Density Correlation Function}

In the case of the density correlation function, $\langle
j(x,t)j(0,0)\rangle $, where $j(x)=\fad(x)\fa(x)$, we have $\Delta
N=0$ so the most general excitation is constructed by backscattering
$d$ particles and creating a particle-hole pair at the Fermi surface
characterized by $N^\pm.$ Making use of
(\ref{backnenergy},\ref{backnmomenta},\ref{jumpn}), we obtain for
the energy and momentum gap of the excitation characterized by
$Q=\{\Delta N=0,d,N^\pm\}$: \be
P_{N^\pm,d}-P_0=2k_Fd+\frac{2\pi}{L}(N^+-N^-)\, , \ee \be
E_{N^\pm,d}-E_0=\frac{2\pi v_F}{L}[(\mathcal{Z}d)^2+N^++N^-]\, . \ee
Here we have taken into account only the terms of order $1$ and
$\mathcal{O}(1/L)$. Equation (\ref{gapcdconn}) gives the conformal
dimensions \be\label{confdjj}
2\Delta^\pm_Q=2N^\pm+(\mathcal{Z}d)^2\, , \ee and from the general
formula (\ref{corrgeneral}) \be \label{denscorgen} \langle
j(x,t)j(0,0) \rangle-\langle
j(0,0)\rangle^2=\sum_{Q=\{N^\pm,d\}}A(Q)\frac{e^{2ixk_Fd}}{
(ix+v_F\tau)^{2\Delta^+_Q}(-ix+v_F\tau)^{2\Delta^-_Q}}\, . \ee

Defining $\theta \equiv 2\mathcal{Z}^2=4\pi D/v_F$, where
$\mathcal{Z}=Z(-q)=Z(q)$, and $Z(\la)$ given by the integral
equation (\ref{zie}), the leading terms are \be \langle j(x,t)j(0,0)
\rangle-\langle
j(0,0)\rangle^2=\frac{a}{(ix+v_F\tau)^2}+\frac{a}{(-ix+v_F\tau)^2}+
b\frac{\cos(2k_Fx)}{|ix+v_F\tau|^\theta}\, . \ee For equal times,
Eq.~(\ref{denscorgen}) takes the form \be\label{denscoret} \langle
j(x,0)j(0,0) \rangle-\langle j(0,0)\rangle^2=\sum_{Q=\{N^\pm,d\}}
\hat{A}(Q)\frac{e^{2ixk_Fd}}{|x|^{d^2\theta+2N^++2N^-}}\, . \ee The
presence of the oscillatory terms in this expression can be
explained by the following simple computation \cite{BM2}:
\begin{eqnarray}
\langle j(x,0)j(0,0)\rangle&=&\sum_Q\langle 0|j(x,0)|Q\rangle\langle
Q|j(0,0)|0\rangle =\sum_Q|\langle0|j(0,0)|Q\rangle|^2e^{i(P_Q-P_0)x}=
\nonumber \\
&=&\sum_{d=-\infty}^\infty e^{i2k_Fdx}\sum_{N^\pm}|\langle
0|j(0,0)|d,N^\pm\rangle|^2e^{\frac{i2\pi x}{L}(N^+-N^-)}\, ,
\end{eqnarray}
where in the second line, we broke the sum over $Q$ into disjoint
sums characterized by different macroscopic momenta. The second part
of the sum gives the power-law decay  for $k_F^{-1}\ll x\ll L$. The
formulae (\ref{denscorgen}) and (\ref{denscoret}) are the same as in
the case of a Bose gas with coupling constant
$c'=c/\cos(\pi\kappa/2)$ and periodic boundary conditions \cite{BM2}
-- see Chap.~XVII of \cite{KBI}. This situation is expected, since
\be j(x)=\fad(x)\fa(x)=\fbd(x)\fb(x)\, . \ee

Finally, from (\ref{cortemp}), the finite temperature density
correlation function is \be \label{denscorgent} \langle
j(x,t)j(0,0)\rangle_T=\sum_{Q=\{d,N^\pm\}}
B(Q)e^{i2k_Fdx}\left(\frac{\pi T/v_F}{\sinh[\pi
T(x-iv_F\tau)/v_F]}\right)^{2\Delta^+_Q} \left(\frac{\pi
T/v_F}{\sinh[\pi T(x+iv_F\tau)/v_F]}\right)^{2\Delta^-_Q}\, , \ee
with $\Delta^\pm_Q$ given by (\ref{confdjj}).

\subsection{Field-Field Correlator}

In contrast to the density correlators, for the field correlator
$\langle \fa(x,t)\fad(0,0) \rangle$, one has $\Delta N=1$. For the
ground states with $N$ and $N+1$ particles and the boundary
conditions considered in this Section the Bethe equations are:
\begin{eqnarray}
\lambda_jL+\sum_{k=1}^N\theta(\lambda_j-\lambda_k)&=&2\pi\left(
j-\frac{N+1}{2}\right)\, , \ \ \ \ \ j=1,\cdots,N\, ,\nonumber\\
\tilde\lambda_jL+\sum_{k=1}^{N+1}\theta(\tilde\lambda_j-
\tilde\lambda_k)&=&2\pi\left(j-\frac{N+2}{2}\right)+\pi\kappa\, , \
\ \ \  j=1,\cdots,N+1\, .
\end{eqnarray}
The shift $\pi \kappa$ in the second equation implies that the
anyonic wavefunctions for $N$ and $N+1$ particles live in two
orthogonal sectors of the Hilbert space. The addition of one
particle produces in this case a macroscopic change in the momentum,
$\pi k_F\kappa+\pi\kappa/L$, which gives rise to oscillations even
in the dominant term of the field correlator.

The most general excitation is obtained by an addition of one
particle to the system, followed by the backscattering of $d$
particles and creation of a particle-hole pair at the Fermi surface.
Using the results
(\ref{1penergy},\ref{1pmomenta},\ref{backnenergy},\ref{backnmomenta},\ref{jumpn})
with $\omega=0,\omega'=\kappa/2$, we obtain the following
expressions for the energy and momentum gaps of an excitation with
$Q=\{\Delta N=1,d,N^\pm\}$ (retaining, as before, the terms of order
$1$ and $\mathcal{O}(1/L)$): \be P_{N^\pm,d}^{\Delta
N=1}-P_0=2k_F(d+\kappa/2)+\frac{2\pi}{L}\left[(d+\kappa/2)+N^+
-N^-\right]\, , \ee \be E_{N^\pm,d}^{\Delta N=1}-E_0=\frac{2\pi
v_F}{L}\left[\left(\frac{1}{2\mathcal{Z}}\right)^2+\mathcal{Z}^2
(d+\kappa/2)^2+N^++N^-\right]\, , \ee so the conformal dimensions
are \be\label{confdff}
2\Delta^\pm_Q=2N^\pm+\left(\frac{1}{2\mathcal{Z}}\pm\mathcal{Z}
(d+\kappa/2)\right)^2\, . \ee From Eq.~(\ref{corrgeneral}), the
field correlator is \be \label{fieldcorgen} \langle
\fa(x,t)\fad(0,0)
\rangle=\sum_{Q=\{N^\pm,d\}}A(Q)\frac{e^{2ik_F(d+\frac{\kappa}{2})x
}}{(ix+v_F\tau)^{-2\Delta^+_Q}(-ix+v_F\tau)^{-2\Delta^-_Q}}\, , \ee
or in the equal-time case \be \langle \fa(x,0)\fad(0,0)
\rangle=\sum_{Q=\{N^\pm,d\}}\hat
A(Q)\frac{e^{2ik_F(d+\frac{\kappa}{2})x}}
{|x|^{(d+\frac{\kappa}{2})^2\theta+\frac{1}{\theta}+2N^++2N^-}}\, ,
\ee where $\theta=2\mathcal{Z}^2.$ Again, we can heuristically
justify the presence of the oscillatory terms in the correlation
function in the same way as for the density correlator, but for the
field correlator, the complete set of states that is inserted
between $\fa$ and $\fad$ is from the sector with $N+1$ particles
\begin{eqnarray}
\langle \fa(x,0)\fad(0,0)\rangle&=&\sum_Q\langle 0|\fa(x,0)|Q\rangle
\langle Q|\fad(0,0)|0\rangle=\sum_Q|\langle0|\fa(0,0)|Q\rangle|^2e^{i(P_Q-P_0)x}=
\nonumber \\
&=&\sum_{d=-\infty}^\infty e^{i2k_F(d+\frac{\kappa}{2})x}
\sum_{N^\pm} |\langle 0|\fa(0,0)|d,N^\pm\rangle|^2e^{\frac{i2\pi
x}{L}(N^+-N^-)}\, .
\end{eqnarray}
In this case, the terms of the correlation function containing
$e^{i2k_F(d+\kappa/2)x}$ that are responsible for the oscillatory
behavior at $x\ll L$, exhibit dependence on the statistical
parameter.

Equation (\ref{fieldcorgen}) can be compared to the result of
Calabrese and Mintchev \cite{CM}, who calculated the field
correlation function for anyonic gapless systems in the low-momentum
regime using the harmonic fluid approach \cite{Hal1,Caz}, obtaining
\be \langle \fad(x,0)\fa(0,0) \rangle=D\sum_{d=-\infty}^\infty
b_d\frac{e^{-2i(d+\frac{\kappa}{2})k_Fx}e^{-2i(m+\frac{\kappa}{2})
\pi\epsilon(x)/2}}{(Dc(x))^{(d+\frac{\kappa}{2})^22K+\frac{1}{2K}}}\,
, \ee where $D$ is the density, $b_d$ unknown non-universal
amplitudes, $c(x)=L\sin(\pi x/L)$, and $K$ is a universal parameter
that can be expressed in terms of the phenomenological velocity
parameters $v_N,v_J$ as $K=\sqrt{v_J/v_N}$. For the Lieb-Liniger
anyons, \be K=\frac{2\pi D}{v_F}=\frac{\theta}{2}\, . \ee They have
checked their results in the limit $c\rightarrow\infty, K=1$ against
the exact results of Santachiara {\it et al.} \cite{SSC}, who
calculated the generalization of Lenard formula \cite{L} for anyonic
statistics. We see that our conformal field theory approach agrees
with the leading asymptotics produced by the harmonic liquid
approximation but also gives the higher-order terms in the
large-distance expansion.

Using the conformal mapping (\ref{tmap}) that leads to general
Eq.~(\ref{cortemp}), we find also the finite-temperature field
correlator:  \be \label{fieldcorgent} \langle
\fa(x,t)\fad(0,0)\rangle_T=\sum_{Q=\{d,N^\pm\}}
B(Q)e^{i2k_F(d+\frac{\kappa}{2})x}\left(\frac{\pi T/v_F}{\sinh[\pi
T(x-iv_F\tau)/v_F]}\right)^{2\Delta^+_Q} \left(\frac{\pi
T/v_F}{\sinh[\pi T(x+iv_F\tau)/v_F]}\right)^{2\Delta^-_Q}\, , \ee
where $\Delta^\pm_Q$ is given by (\ref{confdff}).

\section{Conclusions}

The main result of our work is the calculation of the large-distance
asymptotics of the correlation functions of the gas of 1D anyons
using the ideas of conformal field theory. This result requires
conformal invariance close to the critical point $T=0$, and the
knowledge of the finite size corrections to the energy and momentum
of the ground state of the gas due to low-lying excitations. In the
analogous case of Bose gas with $\delta$-function repulsive
interaction, the conformal field theory predictions for the
asymptotics of the correlators were checked against the exact
results for these asymptotics obtained from the determinant
representations and the differential equations for the correlation
functions \cite{KBI}. It would be interesting to have similar exact
results for the model studied in this paper which is a natural
anyonic extension of the Bose gas. As a first step in this
direction, Santachiara, Stauffer and Cabra \cite{SSC}, already
obtained for the one-particle reduced density matrix (field
correlator) in the impenetrable limit a representation in terms of
the determinant of a Toeplitz matrix of dimension $(N-1)\times(N-1)$
where $N$ is the number of particles. The exact results for the
anyon correlation functions would also be needed to extend the
correlators derived in this work for essentially one type of
boundary conditions to more general quasiperiodic conditions. This
problem seems particularly natural for anyons for which the
effective boundary conditions for quasiparticle momenta change with
the total number of particles in the system.

\acknowledgments

This work was supported in part by the NSF grants  DMR-0325551, DMR-0302758 and DMS-0503712.

\appendix

\section{Boundary Conditions for the Multi-Anyon Wavefunctions}
\label{bcaa}

In this Appendix, we derive the exact form of the cyclic boundary
conditions for the wavefunctions of the many-anyon system. Our
treatment generalizes the approach of \cite{AN} to the case of
several penetrable particles. In physical terms, the situation we
consider corresponds to anyons confined to move along a loop with,
in general, an external phase shift $\phi$ created, e.g., by a
magnetic filed threading the loop. We start with the case of {\em
two particles} and no external phase shift, $\phi=0$. The
Bethe-Anzatz wavefunction (\ref{wf}) reduces in this case to the
following form: In the region I ($x_1<x_2$) one has \be \chi_I
(x_1,x_2)=\frac{e^{i \pi\kappa/2}}{\sqrt{2[ (\lambda_2
-\lambda_1)^2+c'^2]}} \{ e^{i(x_1\la_1+x_2\la_2)}(\la_2-\la_1-ic')
+e^{i(x_1\la_2+x_2\la_1) }(\la_2-\la_1+ic')\}\, , \ee and in the
region II ($x_1>x_2$): \be \chi_{II}(x_1,x_2)=\frac{e^{-i
\pi\kappa/2}}{\sqrt{2[ (\lambda_2-\lambda_1)^2+c'^2]}}\{e^{i
(x_1\la_1 +x_2\la_2)}(\la_2 -\la_1+ic')+e^{i(x_1\la_2+x_2
\la_1)}(\la_2-\la_1-ic')\} \, . \ee The general exchange symmetry of
this wavefunction given by Eq.~(\ref{anyonicproperty}) imply that
for fractional $\kappa$ it can not satisfy the same boundary
conditions in the two coordinates. As one can see by exchanging the
coordinates, if the wavefunction is periodic in the first one, the
boundary conditions in second one should have a twist, \be
\label{bcch} \chi(0,x)=\chi(L,x) \;\;\; \rightarrow \;\;\;
\chi(x,0)=\chi(x,L) e^{-2 i \pi \kappa} \, , \ee and viceversa. One
consequence of this is that the exact form of the Bethe equations
(\ref{BE}) depends on whether we impose periodic boundary conditions
on one or the other coordinate. Indeed, if one requires periodicity
in $x_1$, $\chi(0,x_2)=\chi(L,x_2)$, the Bethe equations are: \be
e^{iL\la_j}=e^{i\pi\kappa}\prod_{k=1,k\ne j}^2 \left(\frac{\la_j
-\la_k +ic'}{\la_j-\la_k-ic'}\right) \, , \ee whereas the
periodicity in $x_2$, $\chi(x_1,0)= \chi(x_1,L)$, results in the
equations that differ by the sign of the statistics parameter
$\kappa$:  \be e^{iL\la_j}= e^{-i\pi \kappa} \prod_{k=1,k\ne
j}^2\left(\frac{\la_j-\la_k+ic'}{ \la_j- \la_k-ic'} \right) \, . \ee
Since the Bethe equations determine the spectrum of the
quasiparticle momenta $\la_j$ through Eq.~(\ref{BEM}), the $\kappa$
shifts of different signs produce two physically different
situations.

The origin of this difference can be traced back to the fact that
the fractional statistics requires braiding of particles, something
that strictly speaking can not be done in one dimension. To define
the braiding of 1D particles one needs to first adopt a convention
on how the particles pass each other at coinciding points, something
that is done by choosing a specific sign of the exchange phase
$e^{-i\pi \kappa \epsilon(x_1-x_2)/2}$. After that, one more choice
that needs to be made is how the 1D loop with anyons is imbedded
into the underlying 2D anyonic system. In the case of two particles,
this choice is reflected in the possibility of choosing different
boundary conditions for two different anyonic coordinates and
determines how the particle trajectories enclose each other as the
particles move along the loop \cite{AN}. As reflected in
Eq.~(\ref{bcch}), periodicity in $x_1$ means that the trajectory of
$x_1$ does not enclose the particle $x_2$. This implies that $x_1$
is itself enclosed by the trajectory of $x_2$, producing the twist
in the boundary condition for $x_2$ variable. The different choice
of the boundary condition would mean that the 1D loop in imbedded
into the 2D system in such a way that the trajectory of $x_1$
encloses $x_2$. This means that the wavefunction periodicity in both
variables correspond to different but valid physical situations.

The situation is somewhat more complicated for larger number of
particles, as can be seen in the case of {\em three particles}. In
the wavefunction (\ref{wf}), one needs to distinguish then six
regions corresponding to the six permutation of the particles. The
wavefunction (\ref{wf}) in these regions is:\\
Region I $(x_1<x_2<x_3)$
\begin{eqnarray}
\chi_I(x_1,x_2,x_3)            &=&Ae^{\frac{i3\pi\kappa}{2}}\left\{ e^{i(x_1\la_1+x_2\la_2+x_3\la_3)}(\la_3-\la_2-ic')(\la_3-\la_1-ic')(\la_2-\la_1-ic')\right. \nonumber\\
                               & &\ \ \ \ \ \ \ \ -e^{i(x_1\la_1+x_2\la_3+x_3\la_2)}(\la_2-\la_3-ic')(\la_2-\la_1-ic')(\la_3-\la_1-ic')\nonumber\\
                               & &\ \ \ \ \ \ \ \ +e^{i(x_1\la_3+x_2\la_1+x_3\la_2)}(\la_2-\la_1-ic')(\la_2-\la_3-ic')(\la_1-\la_3-ic')\nonumber\\
                               & &\ \ \ \ \ \ \ \ -e^{i(x_1\la_3+x_2\la_2+x_3\la_1)}(\la_1-\la_2-ic')(\la_1-\la_3-ic')(\la_2-\la_3-ic') \\
                               & &\ \ \ \ \ \ \ \ +e^{i(x_1\la_2+x_2\la_3+x_3\la_1)}(\la_1-\la_3-ic')(\la_1-\la_2-ic')(\la_3-\la_2-ic')\nonumber\\
                        & &\left.\ \ \ \ \  \ \ \
                        -e^{i(x_1\la_2+x_2\la_1+x_3\la_3)}(\la_3-\la_1-ic')(\la_3-\la_2-ic')(\la_1-\la_2-ic')\right\}
                        ,                        \nonumber
\end{eqnarray}
Region II $(x_1<x_3<x_2)$
\begin{eqnarray}
\chi_{II}(x_1,x_2,x_3)&=&Ae^{\frac{i \pi\kappa}{2}}\left\{ e^{i(x_1\la_1+x_2\la_2+x_3\la_3)}(\la_3-\la_2+ic')(\la_3-\la_1-ic')(\la_2-\la_1-ic')\right.\nonumber\\
                                  & &\ \ \ \ \ \ \ \ -e^{i(x_1\la_1+x_2\la_3+x_3\la_2)}(\la_2-\la_3+ic')(\la_2-\la_1-ic')(\la_3-\la_1-ic')\nonumber\\
                                  & &\ \ \ \ \ \ \ \ +e^{i(x_1\la_3+x_2\la_1+x_3\la_2)}(\la_2-\la_1+ic')(\la_2-\la_3-ic')(\la_1-\la_3-ic')\nonumber\\
                                  & &\ \ \ \ \ \ \ \ -e^{i(x_1\la_3+x_2\la_2+x_3\la_1)}(\la_1-\la_2+ic')(\la_1-\la_3-ic')(\la_2-\la_3-ic') \\
                                  & &\ \ \ \ \ \ \ \ +e^{i(x_1\la_2+x_2\la_3+x_3\la_1)}(\la_1-\la_3+ic')(\la_1-\la_2-ic')(\la_3-\la_2-ic')\nonumber\\
                           & &\left.\ \ \ \ \  \ \ \
                           -e^{i(x_1\la_2+x_2\la_1+x_3\la_3)}(\la_3-\la_1+ic')(\la_3-\la_2-ic')(\la_1-\la_2-ic')\right\} , \nonumber
\end{eqnarray}
Region III $(x_3<x_1<x_2)$
\begin{eqnarray}
\chi_{III}(x_1,x_2,x_3)&=&Ae^{\frac{-i\pi\kappa}{2}}\left\{ e^{i(x_1\la_1+x_2\la_2+x_3\la_3)}(\la_3-\la_2+ic')(\la_3-\la_1+ic')(\la_2-\la_1-ic')\right.\nonumber\\
                                  & &\ \ \ \ \ \ \ \ -e^{i(x_1\la_1+x_2\la_3+x_3\la_2)}(\la_2-\la_3+ic')(\la_2-\la_1+ic')(\la_3-\la_1-ic')\nonumber\\
                                  & &\ \ \ \ \ \ \ \ +e^{i(x_1\la_3+x_2\la_1+x_3\la_2)}(\la_2-\la_1+ic')(\la_2-\la_3+ic')(\la_1-\la_3-ic')\nonumber\\
                                  & &\ \ \ \ \ \ \ \ -e^{i(x_1\la_3+x_2\la_2+x_3\la_1)}(\la_1-\la_2+ic')(\la_1-\la_3+ic')(\la_2-\la_3-ic') \\
                                  & &\ \ \ \ \ \ \ \ +e^{i(x_1\la_2+x_2\la_3+x_3\la_1)}(\la_1-\la_3+ic')(\la_1-\la_2+ic')(\la_3-\la_2-ic')\nonumber\\
                           & &\left.\ \ \ \ \  \ \ \
                           -e^{i(x_1\la_2+x_2\la_1+x_3\la_3)}(\la_3-\la_1+ic')(\la_3-\la_2+ic')(\la_1-\la_2-ic')\right\} , \nonumber
\end{eqnarray}
Region IV $(x_3<x_2<x_1)$
\begin{eqnarray}
\chi_{IV}(x_1,x_2,x_3)&=&Ae^{\frac{-i3\pi\kappa}{2}}\left\{ e^{i(x_1\la_1+x_2\la_2+x_3\la_3)}(\la_3-\la_2+ic')(\la_3-\la_1+ic')(\la_2-\la_1+ic')\right.\nonumber\\
                                   & &\ \ \ \ \ \ \ \ -e^{i(x_1\la_1+x_2\la_3+x_3\la_2)}(\la_2-\la_3+ic')(\la_2-\la_1+ic')(\la_3-\la_1+ic')\nonumber\\
                                   & &\ \ \ \ \ \ \ \ +e^{i(x_1\la_3+x_2\la_1+x_3\la_2)}(\la_2-\la_1+ic')(\la_2-\la_3+ic')(\la_1-\la_3+ic')\nonumber\\
                                   & &\ \ \ \ \ \ \ \ -e^{i(x_1\la_3+x_2\la_2+x_3\la_1)}(\la_1-\la_2+ic')(\la_1-\la_3+ic')(\la_2-\la_3+ic') \\
                                   & &\ \ \ \ \ \ \ \ +e^{i(x_1\la_2+x_2\la_3+x_3\la_1)}(\la_1-\la_3+ic')(\la_1-\la_2+ic')(\la_3-\la_2+ic')\nonumber\\
                            & &\left.\ \ \ \ \  \ \ \ -e^{i(x_1\la_2+x_2\la_1+x_3\la_3)}(\la_3-\la_1+ic')(\la_3-\la_2+ic')(\la_1-\la_2+ic')\right\} ,
                            \nonumber
\end{eqnarray}
Region V $(x_2<x_1<x_3)$
\begin{eqnarray}
\chi_V(x_1,x_2,x_3)&=&Ae^{\frac{i\pi\kappa}{2}}\left\{ e^{i(x_1\la_1+x_2\la_2+x_3\la_3)}(\la_3-\la_2-ic')(\la_3-\la_1-ic')(\la_2-\la_1+ic')\right.\nonumber\\
                                 & &\ \ \ \ \ \ \ \ -e^{i(x_1\la_1+x_2\la_3+x_3\la_2)}(\la_2-\la_3-ic')(\la_2-\la_1-ic')(\la_3-\la_1+ic')\nonumber\\
                                 & &\ \ \ \ \ \ \ \ +e^{i(x_1\la_3+x_2\la_1+x_3\la_2)}(\la_2-\la_1-ic')(\la_2-\la_3-ic')(\la_1-\la_3+ic')\nonumber\\
                                 & &\ \ \ \ \ \ \ \ -e^{i(x_1\la_3+x_2\la_2+x_3\la_1)}(\la_1-\la_2-ic')(\la_1-\la_3-ic')(\la_2-\la_3+ic') \\
                                 & &\ \ \ \ \ \ \ \ +e^{i(x_1\la_2+x_2\la_3+x_3\la_1)}(\la_1-\la_3-ic')(\la_1-\la_2-ic')(\la_3-\la_2+ic')\nonumber\\
                          & &\left.\ \ \ \ \  \ \ \ -e^{i(x_1\la_2+x_2\la_1+x_3\la_3)}(\la_3-\la_1-ic')(\la_3-\la_2-ic')(\la_1-\la_2+ic')\right\} ,
                          \nonumber
\end{eqnarray}
Region VI $(x_2<x_3<x_1)$
\begin{eqnarray}
\chi_{VI}(x_1,x_2,x_3)&=&Ae^{\frac{-i\pi\kappa}{2}}\left\{ e^{i(x_1\la_1+x_2\la_2+x_3\la_3)}(\la_3-\la_2-ic')(\la_3-\la_1+ic')(\la_2-\la_1+ic')\right.\nonumber\\
                                  & &\ \ \ \ \ \ \ \ -e^{i(x_1\la_1+x_2\la_3+x_3\la_2)}(\la_2-\la_3-ic')(\la_2-\la_1+ic')(\la_3-\la_1+ic')\nonumber\\
                                  & &\ \ \ \ \ \ \ \ +e^{i(x_1\la_3+x_2\la_1+x_3\la_2)}(\la_2-\la_1-ic')(\la_2-\la_3+ic')(\la_1-\la_3+ic')\nonumber\\
                                  & &\ \ \ \ \ \ \ \ -e^{i(x_1\la_3+x_2\la_2+x_3\la_1)}(\la_1-\la_2-ic')(\la_1-\la_3+ic')(\la_2-\la_3+ic') \\
                                  & &\ \ \ \ \ \ \ \ +e^{i(x_1\la_2+x_2\la_3+x_3\la_1)}(\la_1-\la_3-ic')(\la_1-\la_2+ic')(\la_3-\la_2+ic')\nonumber\\
                           & &\left.\ \ \ \ \  \ \ \ -e^{i(x_1\la_2+x_2\la_1+x_3\la_3)}(\la_3-\la_1-ic')(\la_3-\la_2+ic')(\la_1-\la_2+ic')\right\},
                           \nonumber
\end{eqnarray}
where \be A=\frac{1}{\sqrt{6 \prod_{j>k}[(\lambda_j-\lambda_k)^2
+c'^2] }} \, . \ee

As discussed above for the two particles, the periodic boundary
conditions can be imposed in principle on any of the wavefunction
arguments. Requiring $x_1$ to be periodic,
$\chi(0,x_2,x_3)=\chi(L,x_2,x_3) $, gives
\begin{eqnarray}
\chi_I(0,x_2,x_3)&=&\chi_{VI}(L,x_2,x_3)\, , \;\;\;  \mbox{ for }
x_2<x_3 \, ,
\label{periodicfirst1} \\
\chi_{II}(0,x_2,x_3)&=&\chi_{IV}(L,x_2,x_3)\, , \;\;\; \mbox{ for }
x_3<x_2\, . \label{periodicfirst2}
\end{eqnarray}
Except for the exchange-statistics phase factors, the wavefunctions
in the six regions coincide with the wavefunctions of the Bose gas
with the $\delta$-function interaction of strength $c'$ (\ref{c}).
Therefore, the Bethe equations we obtain are the same as in the
bosonic case with the only difference coming from the statistical
phase factors. Conditions (\ref{periodicfirst1}) and
(\ref{periodicfirst2}) produce six equations each, with only three
of them being independent
\begin{eqnarray}
e^{iL\la_1}&=&e^{2i\pi\kappa}\left(\frac{\la_1-\la_2+ic'}{\la_1-\la_2-ic'}\right)
\left(\frac{\la_1-\la_3+ic'}{\la_1-\la_3-ic'}\right) , \nonumber \\
e^{iL\la_2}&=&e^{2i\pi\kappa}\left(\frac{\la_2-\la_1+ic'}{\la_2-\la_1-ic'}\right)
\left(\frac{\la_2-\la_3+ic'}{\la_2-\la_3-ic'}\right), \\
e^{iL\la_3}&=&e^{2i\pi\kappa}\left(\frac{\la_3-\la_1+ic'}{\la_3-\la_1-ic'}\right)
\left(\frac{\la_3-\la_2+ic'}{\la_3-\la_2-ic'}\right). \nonumber
\end{eqnarray}
These equations can be written in the compact form similar to
Eq.~(\ref{BE}): \be\label{Bethe1pbc}
e^{iL\la_j}=e^{2i\pi\kappa}\prod_{k=1,k\ne
j}^3\left(\frac{\la_j-\la_k+ic'}{\la_j-\la_k-ic'}\right) . \ee If
the periodic boundary conditions are imposed on the second variable,
$\chi(x_1,0,x_3)=\chi(x_1,L,x_3)$, i.e.,
\begin{eqnarray}
\chi_{V}(0,x_2,x_3)&=&\chi_{II}(L,x_2,x_3)\, , \;\;\; \mbox{ for }
x_1<x_3\, , \label{periodicsecond1} \\
\chi_{VI}(x_1,0,x_3)&=&\chi_{III}(x_1,L,x_3)\, ,\;\;\; \mbox{ for }
x_1>x_3\, , \label{periodicsecond2}
\end{eqnarray}
we obtain either from (\ref{periodicsecond1}) or
(\ref{periodicsecond2}) the following Bethe equations
\be\label{Bethe2pbc} e^{iL\la_j}=\prod_{k=1,k\ne
j}^3\left(\frac{\la_j-\la_k+ic'}{\la_j-\la_k-ic'}\right) . \ee
Finally, if we impose periodic boundary conditions on the third
variable, $\chi(x_1,x_2,0)=\chi(x_1,x_2,L)$, i.e.,
\begin{eqnarray}
\chi_{III}(x_1,x_2,0)=\chi_{I}(x_1,x_2,L)\, , \;\;\;  \mbox{ for } x_1<x_2\, ,
\label{periodicthird1}\\
\chi_{IV}(x_1,x_2,0)=\chi_V(x_1,x_2,L) \, , \;\;\; \mbox{ for }
x_2<x_1 \, , \label{periodicthird2}
\end{eqnarray}
the resulting Bethe equations are \be \label{Bethe3pbc}
e^{iL\la_j}=e^{-2i\pi\kappa}\prod_{k=1,k\ne
j}^3\left(\frac{\la_j-\la_k+ic'}{\la_j-\la_k-ic'}\right) . \ee

The difference between the three forms of the Bethe equations
(\ref{Bethe1pbc}), (\ref{Bethe2pbc}), (\ref{Bethe3pbc}) means that
the periodic boundary conditions imposed on one variable
automatically require the twisted boundary conditions  on the other
variables if one wants to keep the same Bethe equations. Similarly
to the case of two particles, this can also be seen directly from
the anyonic exchange symmetry (\ref{anyonicproperty}) of the
wavefunction. Suppose we set the periodic boundary conditions on the
first variable: \be \label{pbc1} \chi(0,x_2,x_3)=\chi(L,x_2,x_3)\, .
\ee Exchanging then the first two variables on both sides of
Eq.~(\ref{pbc1}) with the help of Eq.~(\ref{anyonicproperty}), we
get the twisted boundary conditions for the second variable: \be
\label{pbc2} \chi(x_2,0,x_3)= \chi(x_2,L,x_3) e^{-2i\pi\kappa}\, .
\ee From (\ref{pbc2}), using again (\ref{anyonicproperty}) we have
\be\label{pbc3} \chi(x_2,x_3,0)=\chi(x_2,x_3,L)e^{-4i\pi\kappa} \, ,
\ee which are the twisted boundary conditions for the third variable
which follow from the periodic conditions on the first. From any of
the boundary conditions  (\ref{pbc1}), (\ref{pbc2}), (\ref{pbc3}) we
obtain the Bethe equations (\ref{Bethe1pbc}).

Similarly, periodic boundary conditions on the second variable give
the following boundary conditions for the three-anyon wavefunction:
\begin{eqnarray}
\chi(0,x_2,x_3)&=&\chi(L,x_2,x_3)e^{2i\pi\kappa} \, ,\nonumber \\
\chi(x_1,0,x_3)&=&\chi(x_1,L,x_2)\, , \label{ex} \\
\chi(x_2,x_3,0)&=&\chi(x_2,x_3,L)e^{-2i\pi\kappa} \, , \nonumber
\end{eqnarray}
and the Bethe equations (\ref{Bethe2pbc}). The same can be done
starting with periodicity in the third variable. As in the case of
two particles, we see that imposing periodic boundary conditions on
the first and the last variables produces the Bethe equations,
(\ref{Bethe1pbc}) and (\ref{Bethe3pbc}), which differ only by the
sign of the statistical parameter $\kappa$. As discussed in detail
for the two particles, this difference corresponds physically to
different imbedding of the 1D loop of anyons into the underlying 2D
system. In the two situations, the number of particles enclosed by
the trajectories of successive particles $x_j$, $j=1,2,...,N$,
either increases from 0 to $N-1$ or decreases from $N-1$ to 0, as
reflected in the corresponding boundary conditions of the
multi-anyon wavefunction. In contrast to this, the requirement of
periodicity of one of the ``internal'' variables (e.g., $x_2$ in the
case of three particles) produces the Bethe equations and boundary
conditions, e.g. (\ref{Bethe2pbc}) and (\ref{ex}), that do not have
this interpretation. They describe the situations with appropriate
non-vanishing external phase shift $\phi \neq 0$, which twists
uniformly the boundary conditions of all the variables. In the main
text of our paper, we use the periodic boundary conditions with
respect to the first variable of the anyonic wavefunction or
introduce the external twist $\phi= \pi \kappa (N-1)$ which removes
the anyonic shift of the quasiparticle momenta. As follows from the
discussion in this Appendix, the boundary conditions for the
wavefunction of $N$ anyons are given in these two situations by
Eqs.~(\ref{boundarycond}).

\section{Particle-Hole Excitation}\label{pheappendix}

In this appendix we find the energy and momentum of particle-hole
excitations of the gas of anyons. As discussed in the main text, for
twisted boundary conditions ($\beta=1$), the ground state of anyons
is equivalent to that of the Bose gas with periodic boundary
conditions and coupling constant $c'$, so the excitation energy and
momentum coincide in this case with those known for the Bose gas
(see Chap.~I.4 of \cite{KBI}). For periodic boundary conditions
($\beta=0$), the Bethe equations are the same as for the Bose gas
with the boundary conditions twisted by the phase shift
$2\pi\delta$, where $\delta=\{[\pi\kappa(N-1)]\}$. In the case of
one hole with momentum $\lambda_h$  and one particle with momentum
$\lambda_p$ the equations for the ground state and the excited state
are: \be\label{gsph} \mbox{ Ground State, PBC}:\ \ \ \ \ \ \lambda_j
L+\sum_{k=1}^N\theta(\lambda_j-\lambda_k)=2\pi\left(j-\frac{N+1}{2}
\right)+\sh\, , \ \  \ j=1,\cdots,N\, , \ee \be\label{esph} \mbox{
Excited State, PBC}:\ \ \ \ \ \ \tilde\lambda_j
L+\sum_{k=1}^N\theta(\tilde\lambda_j-\tilde\lambda_k)+\theta(
\tilde\lambda_j-\lambda_p)-\theta(\tilde\lambda_j-\lambda_h)=
2\pi\left(j-\frac{N+1}{2}\right)+\sh\, , \  \ j=1,\cdots,N\, . \ee
Comparing the equations for a particle-hole excitation in the case
of twisted boundary conditions \be \mbox{ Ground State, TBC}:\ \ \ \
\ \ \lambda_j^B
L+\sum_{k=1}^N\theta(\lambda_j^B-\lambda_k^B)=2\pi\left(j-
\frac{N+1}{2}\right) , \  \ j=1,\cdots,N\, , \ee \be \mbox{ Excited
State, TBC}:\ \ \ \ \ \ \tilde\lambda_j^B
L+\sum_{k=1}^N\theta(\tilde\lambda_j^B-\tilde\lambda_k^B)+\theta(\tilde
\lambda_j^B-\lambda_p^B)
-\theta(\tilde\lambda_j^B-\lambda_h^B)=2\pi\left(j-\frac{N+1}{2}
\right) , \  \ j=1,\cdots,N\, , \ee with (\ref{gsph}) and
(\ref{esph}), we find the following relations \be
\lambda_j=\lambda_j^B+2\pi\delta/L,\ \ \ \tilde\lambda_j
=\tilde\lambda_j^B+2\pi\delta/L,\ \ \ \ (j=1,\cdots,N) \ee
\be\label{conn} \lambda_p=\lambda_p^B+2\pi\delta/L,\ \ \ \
\lambda_h=\lambda_h^B+2\pi\delta/L\, . \ee

The energy and momentum of this excited state with respect to the
ground state is ($\varepsilon_0(\la)=\la^2-h$):
\begin{eqnarray}\label{phem}
\Delta E(\lambda_p,\lambda_h)&=&\varepsilon_0(\lambda_p)-
\varepsilon_0(\lambda_h)+\sum_{j=1}^N(\varepsilon_0(\tilde\lambda_j)
-\varepsilon_0(\lambda_j))\nonumber\\
                             &=&\varepsilon_0(\lambda_p^B)-
\varepsilon_0(\lambda_h^B)+\sum_{j=1}^N(\varepsilon_0(\tilde\lambda_j^B)
-\varepsilon_0(\lambda_j^B)) +2\frac{\sh}{L}\left(\lambda_p^B-
\lambda_h^B+\sum_{j=1}^N(\tilde\lambda_j^B-\lambda_j^B)\right)\nonumber\\
                             &=&\Delta E^B(\lambda_p^B,\lambda_h^B)+
2\frac{\sh}{L}\Delta P^P(\lambda_h^B,\lambda_p^B)\, ,\\
\Delta P(\lambda_p,\lambda_h)&=&\Delta
P^B(\lambda_p^B,\lambda_h^B)\, ,
\end{eqnarray}
where $\Delta E^B(\lambda_p^B,\lambda_h^B)$ and $\Delta
P^B(\lambda_p^B,\lambda_h^B)$ are the energy and momentum of a
particle-hole excitation in the Bose gas with periodic boundary
conditions, and $\lambda_h^B$ and $\lambda_p^B$ are given by
(\ref{conn}).

From (\ref{phem}) we see that in the case of twisted boundary
conditions, the Fermi velocity $v_F^{TBC}$ will be the same as in
the Bose gas, whereas for the periodic boundary conditions the Fermi
velocity will be modified as \be
v_F^{PBC}=v_F^{TBC}+\frac{4\pi\delta}{L}\, . \ee



\begin{thebibliography}{99}



\bibitem{LM}     J.M. Leinaas and J. Myrheim: Nuovo Cimento B {\bf
37} (1977), 1.
\bibitem{GMS}    G.A. Goldin, R.Menikoff and D.H. Sharp: J.\ Math.\
Phys. {\bf 21} (1980), 650; {\bf 22} (1981), 1664.
\bibitem{FW}     F. Wilczek: Phys.\ Rev.\ Lett. {\bf 49} (1982),
957.
\bibitem{ASW}    D. Arovas, J.R. Schrieffer, and F. Wilczek: Phys.\
Rev.\ Lett. {\bf 53} (1984), 722.
\bibitem{GS}     V.J. Goldman and B. Su: Science {\bf 267}, 1010
(1995).
\bibitem{AN}     D.V. Averin and J.A. Nesteroff: Phys.\ Rev.\ Lett.
{\bf 99} (2007); [{\tt arXiv:0704.0439}].
\bibitem{AG}     D.V. Averin and V.J. Goldman: Solid State Commun.
{\bf 121} (2002), 25.
\bibitem{SFN}    S. Das Sarma, M. Freedman, and C. Nayak: Phys.\
Rev.\ Lett. {\bf 94}  (2005), 166802.
\bibitem{AK}     A. Kitaev: Ann.\ Phys. {\bf 303} (2003), 2.
\bibitem{SJBR}   S.J.B. Rabello, Phys.\ Rev.\ Lett. {\bf 76} (1996),
4007.
\bibitem{AGRPS}  U. Aglietti, L. Griguolo, R. Jackiw, S.Y. Pi, and D.
Seminara, Phys.\ Rev.\ Lett. {\bf 77} (1996), 4406.
\bibitem{FDMH}    F.D.M Haldane, Phys.\ Rev.\ Lett. {\bf 67} (1991),
937.
\bibitem{Kundu}  A. Kundu: Phys.\ Rev.\ Lett. {\bf 83}  (1999), 1275.
\bibitem{BGO}    M.T. Batchelor, X.-W. Guan and N. Oelkers: Phys.\ Rev.\ Lett. {\bf 96}
(2006), 210402; [{\tt cond-mat/0603643}].
\bibitem{BG}     M.T. Batchelor and X.-W. Guan: Phys.\ Rev. {\bf  B 74} (2006), 195121;
[{\tt cond-mat/0606353}].
\bibitem{BGH}    M.T. Batchelor, X.-W. Guan and J.-S. He: J.\ Stat.\ Mech. (2007)
P03007; [{\tt cond-mat/0611450}].
\bibitem{LL}     E.H. Lieb and W. Liniger: Phys.\ Rev. {\bf 130} (1963),
1605.
\bibitem{KBI}    V.E. Korepin, N.M. Bogoliubov, and A.G. Izergin, {\it Quantum
Inverse Scattering Method and Correlation Functions} Cambridge
Univ.\ Press, 1993.
\bibitem{CM}     P.Calabrese and M. Mintchev: [{\tt
cond-mat/0703117}].

\bibitem{LMP}    A. Liguori, M. Mintchev and L. Pilo: Nucl. Phys. {\bf B 569}
(2000), 577.
\bibitem{IT}    N.Ilieva and W. Thirring: Eur.\ Phys.\ J. {\bf C6}
(1999), 705; Theor.\ Mat.\ Phys. {\bf 121} (1999), 1294.
\bibitem{Gir}    M.D. Girardeau: Phys.\ Rev.\ Lett. {\bf 97} (2006),
100402.
\bibitem{Mcguire}J.B. McGuire: J.\ Math.\ Phys. {\bf 5} (1964), 622.
\bibitem{BIR}    N.M. Bogoliubov, A.G. Izergin, and N. Yu. Reshetikhin: JETP Lett.{\bf 44} (1986), 521.
\bibitem{BIR1}   N.M. Bogoliubov, A.G. Izergin, and N. Yu. Reshetikhin: J.\ Phys. {\bf A 20} (1987), 5361.

\bibitem{BM1}    A. Berkovich and G. Murthy: J. Phys. {\bf A 21} (1988), L 395.
\bibitem{BM2}    A. Berkovich and G. Murthy: J. Phys. {\bf A 21} (1988), 3703.
\bibitem{WET}    F. Woynarovich, H.P. Eckle and T.T. Truong: J. Phys. {\bf A 22} (1989), 4027.
\bibitem{BPZ}    A.A. Belavin, A.M. Polyakov and A.B. Zamolodchikov: Nucl. Phys. {\bf B241} (1984), 333.
\bibitem{ISZ}    C. Itzykson, M. Saleur and J.B. Zuber: {\it Conformal Invariance and Applications to Statistical Mechanics }, Singapore:World Scientific, 1986
\bibitem{Cardy1} J.L. Cardy: Lectures given at Les Houches, Session XLIX, 1988 {\it Champs, Cordes et Ph\'{e}nom\`{e}nes Critiques:
                             Fields, Strings and Critical Phenomena}, edt. E. Brezin and J.Zinn-Justin, Elsevier Science Publishers BV, 1989.
\bibitem{Ginsparg}P. Ginsparg: Lectures given at Les Houches, Session XLIX, 1988 {\it Champs, Cordes et Ph\'{e}nom\`{e}nes Critiques:
                             Fields, Strings and Critical Phenomena}, edt. E. Brezin and J.Zinn-Justin, Elsevier Science Publishers BV, 1989.
\bibitem{BCN}    H.W. Bl\"{o}te, J.L. Cardy and M.P. Nightingale: Phys. Rev. Lett. {\bf 56} (1986), 742.
\bibitem{Affleck}I. Affleck: Phys. Rev. Lett. {\bf 56} (1986), 746.
\bibitem{FQS}    D. Friedan, Z. Qui and S. Shenker: Phys. Rev. Lett. {\bf 52} (1984), 1575
\bibitem{DVV}    R. Dijkgraaf, E. Verlinde and H. Verlinde: Comm. Math. Phys. {\bf 115} (1988), 649.
\bibitem{Cardy2} J.L. Cardy: Nucl. Phys. {\bf B270} (1986), 186.
\bibitem{Hal1}   F.D.M. Haldane: Phys. Rev. Lett. {\bf 97} (1981), 1840; J. Phys. {\bf C 15} (1981), 2585.
\bibitem{Caz}    M.A. Cazalilla: J. Phys. {\bf B 37} (2004), S1.
\bibitem{SSC}    R. Santachiara, F. Stauffer and D.C. Cabra: [{\tt cond-mat/0610402}].
\bibitem{L}      A. Lenard: J. Math. Phys. {\bf 5} (1964) 930.


\end{thebibliography}
\end{document}